\documentclass[prl,twocolumn,amsmath,amssymb,superscriptaddress]{revtex4-2}

\usepackage{graphicx}
\usepackage{dcolumn}
\usepackage{bm}
\usepackage{color}
\usepackage{braket}

\begin{document}

\title{Charge Density Wave bending observed by Xfel source acting as a tunable electronic lens for hard x-rays}

\author{E. Bellec}
\affiliation{CEA Grenoble, IRIG, MEM, NRS, 17 rue des Martyrs, F-38000 Grenoble, France}
\author{D. Ghoneim}
\affiliation{Laboratoire de Physique des Solides, Université Paris-Saclay, CNRS, 91405 Orsay, France}
\affiliation{European X-Ray Free-Electron Laser Facility, D-22869 Schenefeld, Germany}
\author{A. Gallo}
\affiliation{Laboratoire de Physique des Solides, Université Paris-Saclay, CNRS, 91405 Orsay, France}
\author{V.L.R. Jacques}
\affiliation{Laboratoire de Physique des Solides, Université Paris-Saclay, CNRS, 91405 Orsay, France}
\author{I. Gonzalez-Vallejo}
\affiliation{Max-Born-Institut fur Nichtlineare Optik und Kurzzeitspektroskopie, 12489 Berlin, Germany}
\author{L. Ortega}
\affiliation{Laboratoire de Physique des Solides, Université Paris-Saclay, CNRS, 91405 Orsay, France}
\author{M. Chollet}
\affiliation{Linac Coherent Light Source, SLAC National Accelerator Laboratory, 2575 Sand Hill Road, Menlo Park, CA 94025, USA}
\author{A. Sinchenko}
\affiliation{Laboratoire de Physique des Solides, Université Paris-Saclay, CNRS, 91405 Orsay, France}
\author{D. Le Bolloc'h}
\affiliation{Laboratoire de Physique des Solides, Université Paris-Saclay, CNRS, 91405 Orsay, France}
\date{\today}

\begin{abstract}
Ultrafast X-ray diffraction by the LCLS free-electron laser has been used to probe Charge Density Wave (CDW) systems under applied external currents.  At sufficiently low currents, CDW wavefronts bend in the direction transverse to the 2k$_F$ wave vector. We show that this shear effect has the ability to focus or defocus hard X-ray beams, depending of the current direction, making it an electronic lens of a new kind, tunable at will from the Fraunhofer to the Fresnel regime. The effect is interpreted using the fractional Fourier transform showing how the macroscopic curvature of a nanometric modulation can be beneficially used to modify the propagation of X-ray beams.
\end{abstract}

\pacs{}

\maketitle


\section{Introduction}
A incommensurate Charge Density Wave (CDW) may be highly sensitive to even small external excitation. It can then appear independent with respect to the atomic host lattice that yet supports it. For example when NbSe$_3$ is cooled down, the atomic lattice contracts slightly while the CDW's period increases considerably \cite{Hodeau1978,mouddenPhysRevLett.65.223}.  In the case of rare-earth tritellurides, the CDW can even flip 90° after an ultra short femto-laser pulse, and return to the equilibrium position in less than few ps \cite{Kogar_naturephys2020, Isabel_struct_dyn2022}. The CDW can also tilt in the perpendicular direction  after elongating the lattice parameters by less than 1$\%$ by using a tensile machine \cite{fisherPhysRevX.12.021046,Gallo2023}. 

A CDW is also very sensitive to external currents. Indeed, when a small external current is applied, the CDW is submitted to a force due to the applied field because the CDW is naturally charged. Two deformations are then observed. The first which concerns the longitudinal direction along the 2k$_F$ wave vector, compresses the CDW near one electrode and expands it at the other \cite{PhysRevLett.70.845,PhysRevLett.80.5631}. The second deformation, more recently observed, is located far from the electrodes. It involves the transverse direction and corresponds to a curvature of the CDW wavefronts \cite{bellec2019evidence,bellec2020}. CDW's defects are described by considering only the phase of the modulation $\eta\cos(2k_F x+\Phi(\vec r))$\cite{lebolloch_mdpi2023}. In this specific case, involving a continuous deformation, contraction, dilatation and shear, a quadratic phase $\Phi(\vec r)$ has to be considered, depending of the applied force $f$, either along the longitudinal direction ($\Phi(x)=f(x^2-l^2_x)$) or the transverse one ($\Phi(y)=f(y^2-l^2_y)$) where $2l_x$ and $2l_y$ are the sample dimensions in the two perpendicular directions.

The transverse shear under current has been observed in NbSe$_3$ by X-ray micro-diffraction. Fast mapping techniques were used to get the quadratic CDW's phase by phase gradient method. The observed curvature of CDW wavefronts increases with rising currents and reverses with inverse currents, until a current threshold is reached, beyond which relaxation is observed. The CDW in NbSe$_3$ is then able to bend continuously over more than 20$\mu m$, that is a distance 10$^4$ larger than its $\lambda$=14$\AA$ period. This phenomenon illustrates the extent to which a CDW is able to deform continuously from one edge to the other edge over macroscopic distances, despite the presence of numerous defects. This effect can be understood by CDW pinning on sample surfaces \cite{bellec2020}.  

In this letter, we have undertaken to observe this same phenomenon, but with a single illumination, from a beam wide enough and fully coherent delivered by the LCLS XFEL source to see the curvature as a whole. We show in the following that the CDW bending induced by the small external current modifies the beam propagation from the Fresnel to the Fraunhofer regime, allowing to continuously focus  hard x-ray beams.

\begin{figure}[b]
\includegraphics[scale= 0.45]{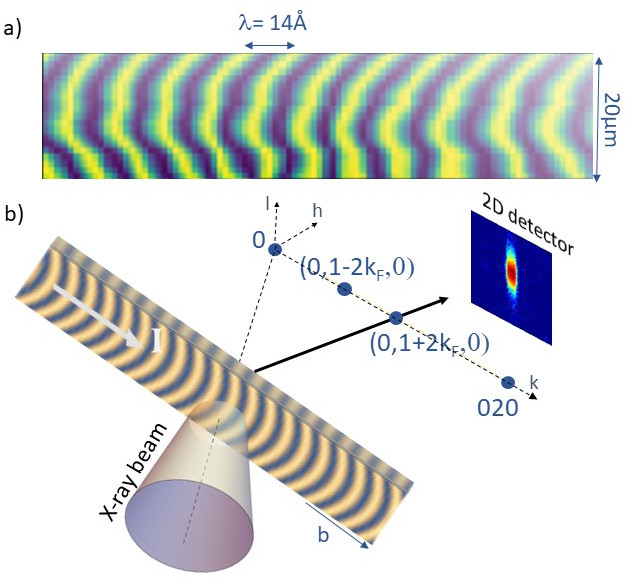}
\caption{a) Curvature of the CDW wavefronts observed in NbSe$_3$  when the sample is submitted to a current of I=-1mA. This image has been obtained from a fast mapping diffraction experiment using a micrometric x-ray beam (from \cite{bellec2019evidence}). For better visibility, the CDW wavelength $\lambda$ has been considerably increased to match the sample size. b) Experimental configuration s illustrating the wide angles diffraction setup.}
\end{figure}
\begin{figure}[b]
\includegraphics[scale= 0.25]{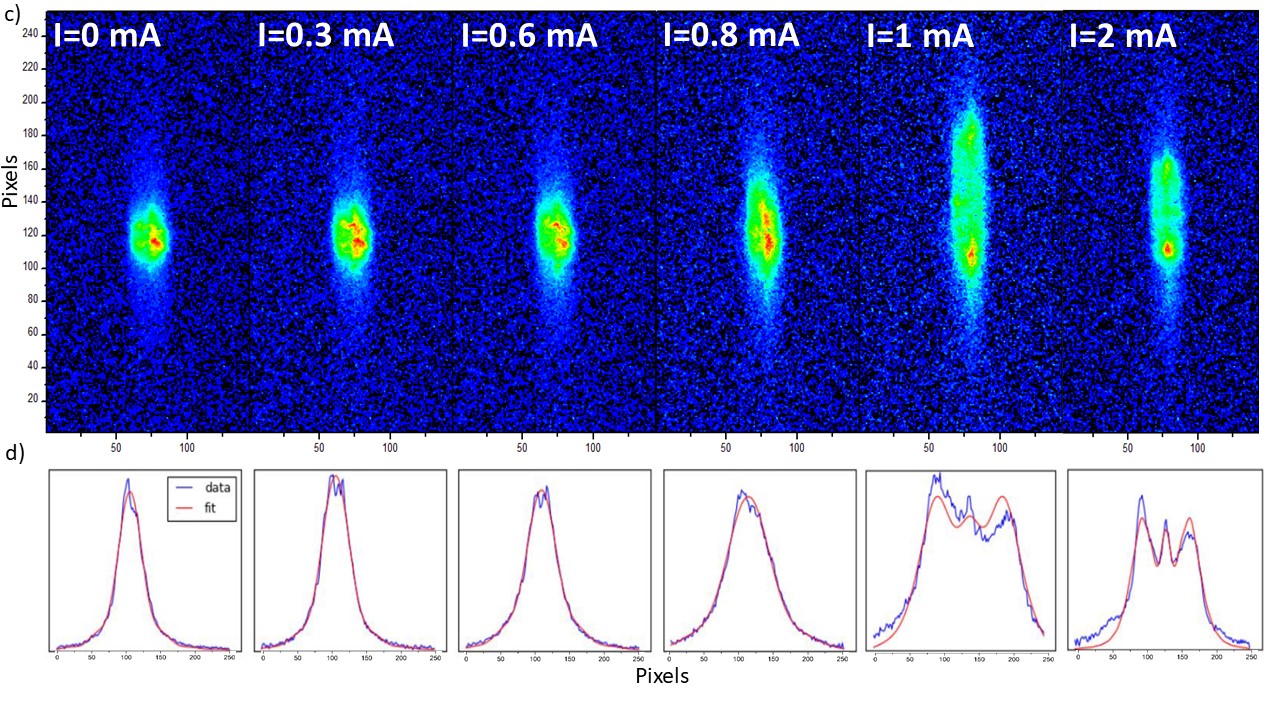}
\caption{c) (0 1.241 0) satellite reflections associated with the CDW in NbSe$_3$ with increasing currents.
All figures correspond to the integrated intensity over the  complete rocking curve. d) Transverse profiles after vertical integration (bleue curves) as well as the fit (red curves) based on the model described in the text.}
\label{figure1}
\end{figure}
Femtosecond X-ray diffraction experiment has thus been  performed to probe the incommensurate NbSe$_3$ system submitted to several $dc$ currents. The two electrical contacts needed for {\it in situ} resistivity measurements were located at more than 500$\mu m$ from each other along the 39$\mu m$ $\times$ 3$\mu m$ $\times$2.25$mm$ single crystal, itself glued on a sapphire substrate. The connected sample was inserted into a cryostat mounted on the XCS diffractometer and cooled down to 80K, below the first CDW transition (Tc$_1 $=145K). The satellite reflections associated with the CDW are located at $\pm\vec{q}_{2kF}$ from each Bragg reflection with the incommensurate wave vector $q_{2kF} = 0.243\pm 0.001 (\times \frac {2\pi} {b} $). The threshold current, measured during the experiment, was equal to I$_s$ = 0.8 mA. Each 8 keV x-ray pulses, of less than 50fs duration, were focused on a spot of about 30$\mu m$ in diameter on the sample, that is slightly smaller than the sample width. The diffracted intensity was recorded with a 55$\mu$m$\times$55$\mu$m pixel size detector located 8m further in the horizontal plane (see figure 1b).
The Bragg peaks reflect properties of the host atomic lattice, while the satellite reflections associated with CDW contain information about both the host crystal lattice and the CDW modulation. Therefore,
rocking curves were made for the $Q_{020}$ Bragg and the (0 $1+\vec {q}_{2kF}$ 0) satellite reflection for several currents, up to I$_s$. 
\begin{figure}[b]
\includegraphics[scale= .3]{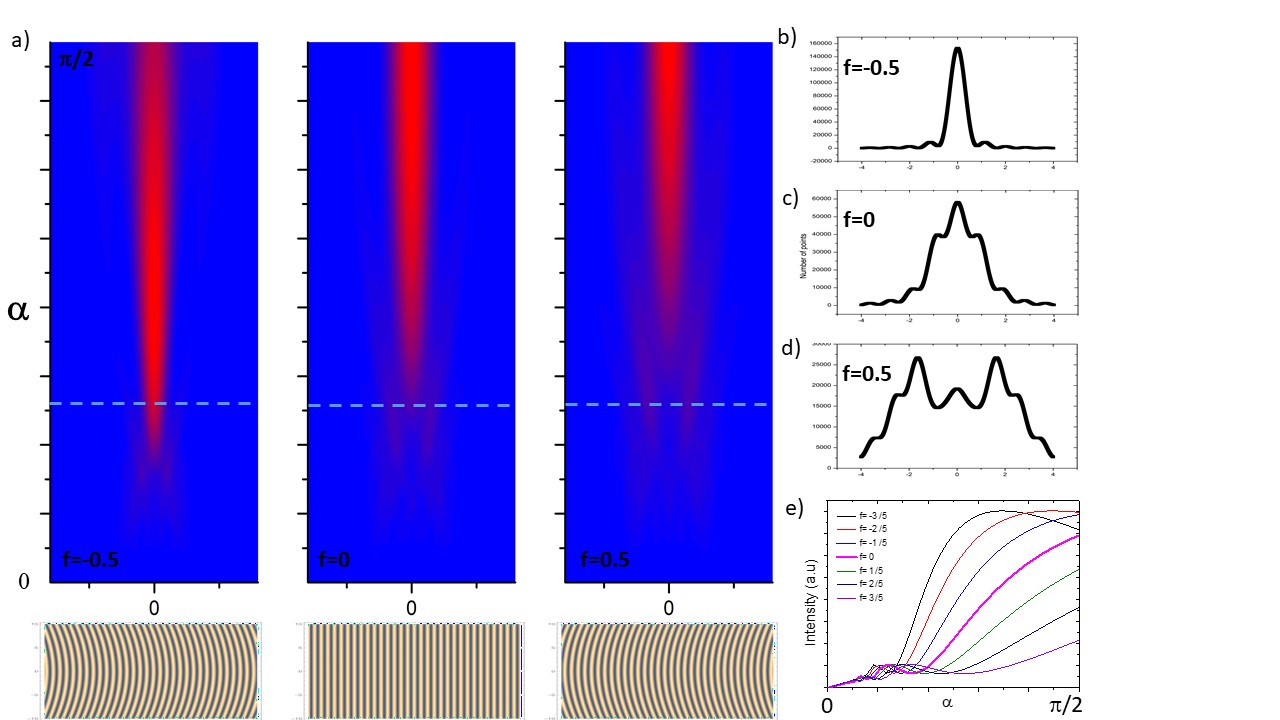}
\caption{a) Slit diffraction from the Fresnel to the Fraunhofer regime calculated by the FrFT for $\alpha\in[0,\pi/2]$ and 2 different forces of oposite signs and below, the corresponding real space showing the respective CDW curvatures. b)-d) Corresponding beam profiles for the 3 forces calculated at a fixed $\alpha$ at the dotted line position. e) Intensity profile at q=0 versus $\alpha$ and forces $f$. The Fresnel-Frauhofer transition is shifted to smaller or larger $\alpha$ values or, equivalently, to smaller or larger distances from the sample by changing the sign of the current applied to the sample.}
\label{figure2}
\end{figure}
The combined effect of the applied electric field and surface pinning induces curvature of CDW wavefronts. As shown in Fig.2, this shear effect has a significant incidence on the diffraction pattern. As currents increase, the beam broadens considerably in the transverse direction and displays characteristic profiles that can easily be explained by diffraction from curved CDW wavefronts.  The diffraction profile of the 2k$_F$ satellite obtained by the large and fully coherent LCLS xfel beam is then obtained from the Fourier Transform of a whole curved modulation, whose exact solution is given in 1D by:
\begin{equation}
\begin{split}
I(q)\propto &(Erfi[\frac{(1+i)(2af-q)}{2\sqrt(2f)}]-Erfi[\frac{(1+i)(2bf-q)}{2\sqrt(2f)}]) \\
&\circledast Gauss(q,\sigma)
\end{split}
\label{erfi}
 \end{equation}
considering that the x-ray beam less wide than the sample width, extending from $a$ to $b$, and a incident Gaussian beam of width $\sigma$. The fit is shown in Fig.2 , which reproduces the main characteristics of the diffraction profiles obtained for different currents by changing only the curvature.

The expression (\ref{erfi}) of the Fourier Transform (FT) of a modulation including a quadratic phase term is exact but not explicit. On the other hand the use of the Fractional Fourier Transform (FrTF) is a more elegant approach to dealing with space curvature and light propagation.  This approach is based on the fact that since the standard Fourier transform must be applied 4 times successively to recover the initial object, the FT is nothing else but an angular transformation of $\alpha=\pi/2$ in the (x, q) plane. A generalized Fourier transform can thus be developed for any angle of rotation $\alpha$ from 0 to $\pi/2$, covering the complete propagation of light from the Fresnel  to the Fraunhofer regime. 
 The integral form of the fractional Fourier transform  of any function $f(x)$ can be written in 1D as \cite{Namias1980}:
 \begin{equation}
 \mathcal{F}_\alpha [g](q)=\int_{-\infty}^\infty K(q,x) g(x) du
 \label{frft}
 \end{equation}
 with
 $$
 K(q,x)=\sqrt{\frac{1-i\cot\alpha}{2\pi}}\exp[{-i\frac{qx}{\sin\alpha}+\frac i2(q^2+x^2)\cot\alpha}]
 $$
The expression of the standard
 Fourier transform is easily recovered for $\alpha=\pi/2$. 

The FrFT is a generalization of the
ordinary FT based on a rotational angle in the usual phase space with applications in optics\cite{Lohmann1993} and x-ray diffraction\cite{lebolloch2011}.
Let's applied the FrFT to the diffraction  of rectangular slit by considering a quadratic phase along the transverse direction $y$. The 1D case reads:
\begin{equation}
\mathcal{F}_\alpha^f (q)\propto\int_{-a}^a K(q,y)\exp[i 2k_F y]\exp[i f(y^2-a^2)]dx 
 \end{equation}
where $f=qE$ is the force induced by the applied current and $\pm a$ the two sample boundaries 
(or the beam size). The effect of the finite and imperfect lattice and the incident beam profile on the diffraction pattern act like a convolution in the Fraunhofer regime and are neglected here.
 
The advantage of using FrFT to deal with space curvature then becomes obvious, since two quadratic terms are present in the integral form. As a consequence, the curvature of the CDW wavefronts under an applied force is equivalent to changing the $ \alpha$ angle with the force $f$ into another $\alpha'$ without force, $\mathcal{F}^f_\alpha(q)=\mathcal{F}^0_{\alpha\prime}(q^\prime)$ with:
$$\cot\alpha'=\cot\alpha+2f$$
and  to rescale in $q$ with the scale factor:
$$q'/q=\frac{1}{\sqrt{1+(\cot\alpha+2f)^2}}$$
We show in Fig.3,  the consequence on the diffraction pattern. The slit diffraction from the Fresnel to the Fraunhofer regime calculated by the FrFT is shown without force $f$. The two regimes are easily accessible from slits closed at few tens of micrometers using keV x-ray beams. By changing $f$, the x-ray propagation is shifted in time, either dilated or contracted depending of the sign of $f$, i.e. depending of the sign of the CDW curvature. As a consequence, if the detector is located in between the two regimes, between Fresnel and Fraunhofer, the diffracted beam will expand and display a characteristic Fresnel profile for positive currents. On the contrary, for negative one, the beam profile will approach the Fraunhofer regime depending of the amplitude of the curvature. It will then be focussed, displaying a cardinal sinus square profile (see Fig.2). 

Two things are worth noting. First, when $\alpha=\pi/2$, i.e. when the detector is located in the Fraunhofer regime, the sign of the curvature no longer plays any role in the diffraction profile: positive and negative currents, of the same intensity, will lead to the same profile, approaching the Fresnel regime. Second, changing the order of diffraction, measuring (0 $1+\vec {q}_{2kF}$ 0) rather than (0 $1+\vec {q}_{2kF}$ 0) reflection, is equivalent to inverse the sign of the curvature.

Focusing beams in the hard x-ray regime is not an easy task and has always been a major limitation in the field of x-rays, especially for large instruments (synchrotron and XFel sources). To focus high energy x-ray beams is now achievable either by using reflective (Kirkpatrick–Baez systems, capillaries, waveguides\cite{Salditt2020}), refractive (Be lenses\cite{snigirev1996}) and diffractive (Fresnel zone plate\cite{NIEMANN1974160}) optics. This work opens up a new class of focusing optics, thanks to the ability of a CDW to bend when subjected to a low current. The resulting nanometric modulation, bent over tens of micrometers, then acts as a versatile electronic lens, tunable at will and operating in a wide energy range. Since the direction of curvature and its amplitude can be controlled by the sign and intensity of the applied current, the diffracted beam can be focused or defocused, from the Fraunhofer to the Fresnel regime. However, this study remains a demonstration of principle because the intensity of the satellite reflection is probably too weak in this compound, but other systems displaying similar properties could prove more effective.

\begin{acknowledgements}
The authors would like to thank G. Abramovici for his help and F. Polack for his always pertinent optician's comments.
\end{acknowledgements}

\bibliography{biblio_lcls}

\end{document}